\def\bea{\begin{eqnarray}}\def\eea{\end{eqnarray}}\def\beq{\begin{equation}} 
\def\eeq{\end{equation}} \def\bed{\begin{displaymath}} 
\def\eed{\end{displaymath}} \def\beqq{\begin{eqnarray}} 
\def\eeqq{\end{eqnarray}} \def\bedd{\begin{eqnarray*}} 
\def\eedd{\end{eqnarray*}} \def\nn{\nonumber} 
\def\be{\begin{equation}}\def\ee{\end{equation}}

\newcommand{\eqngrlb}[3]{\par\parbox{11cm} {\begin{eqnarray}
\fbox{$\displaystyle#1\\#2$}\end{eqnarray}}\hfill 
\parbox{1cm}{\begin{eqnarray}\label{#3}\end{\eqnarray}}\break}  
    \newcommand{\refs}[1]{(\ref{#1})}  
\documentstyle[sprocl]{article}  

\begin{document} 
\bibliographystyle{unsrt} 
\def\pr{\prime} \def\pa{\partial} \def\es{\!=\!} \def\ha{{1\over 2}} 
\def\>{\rangle} \def\<{\langle} \def\mtx#1{\quad\hbox{{#1}}\quad} 
\def\pan{\par\noindent} \def\lam{\lambda} \def\La{\Lambda} 
\def\st{\scriptstyle}
\def\sst{\scriptscriptstyle}
\def\tr{\hbox{tr}} \def\al{\alpha} \def\d{\hbox{d}} \def\De{\Delta} 
\def\L{{\cal L}} \def\H{{\cal H}} \def\Tr{\hbox{Tr}} 
\arraycolsep1.5pt
\def\Journal#1#2#3#4{{#1} {\bf #2}, #3 (#4)}
\def\NCA{\em Nuovo Cimento}\def\NIM{\em Nucl. Instrum. Methods}
\def\NIMA{{\em Nucl. Instrum. Methods} A}\def\NPB{{\em Nucl. Phys.} B}
\def\PLB{{\em Phys. Lett.}  B}\def\PRL{\em Phys. Rev. Lett.}
\def\PRD{{\em Phys. Rev.} D}\def\ZPC{{\em Z. Phys.} C}
\title{On Universality in Black Hole 
Thermodynamics\footnote{Talk given at PASCOS98 based on work in 
collaboration with D. Birmingham, 
R. Emparan and S. Sen.}}
\author{Ivo Sachs}
\address{Department of Mathematical Sciences\\University of Durham\\
Science Site, Durham DH1 3LE, UK}
\maketitle
\abstracts{The low energy decay rates of four- and five dimensional dyonic 
black holes in string theory are equivalently described in terms of an 
effective near horizon $AdS_3$ (BTZ) black hole. It is then argued that 
$AdS_3$ gravity provides an universal microscopic description of the low 
energy dynamics these black holes.}

\section{Introduction}The last three years have seen dramatic progress in the 
understanding of the microscopic description of black holes initiated by the 
identification of certain dyonic black holes and Bogomol'nyi saturated states 
in string theory \cite{Sen,9512031}. Such configuration arise as vacua in 
compactified superstring theory. The microscopic description then requires 
some understanding of the corresponding world sheet conformal field theory  
\cite{9512031}. This was first achieved in \cite{9601029} for the 
$5$-dimensional dyonic black hole. 

Non extremal as well as four dimensional black holes where 
considered in \cite{9602051} and \cite{9603060} respectively. For 
non-extremal black holes they decay rates also agree with an effective string 
theory prediction for a large range of parameters 
\cite{9605234}. There the microscopic derivations are 
less rigorous. However, the effective string theory predictions work much 
better than they should, i.e. produce the correct result for parameters of 
the black hole where higher order effects on string theory side are 
expected \cite{9609026}. This hints to the existence of a simpler microscopic 
description of these black holes that does not involve all degrees of freedom 
of superstring theory.

Here we elaborate on these ideas. It has been known for some time \cite{Hyun} 
that the near horizon geometry of the five- and six dimensional black string 
is a $AdS_3$ (BTZ \cite{BTZ}) black hole $\times S^{2(3)}$ 
respectively. Furthermore, the s-wave decay rates of the four- and five 
dimensional black holes are identical with the $BTZ$ decay rates 
\cite{BSS1}. On the other hand $AdS_3$ gravity has an equivalent description 
in terms of a conformal field theory at the boundary \cite{Brown86,Carlip}. 
The scalar field induces an interaction between the left- and right moving 
sectors of this $CFT$ and the corresponding normalised transition amplitudes 
indeed reproduce the correct semiclassical $s$-wave decay rates. 

\section{Effective 2+1 dimensional description} 
The (generalised) five- dimensional Reissner Nordstr\"om black hole is 
obtained by compactification of the $6$-dimensional black string with 
metric given by 
\bea \label{5bs}
\d s^2&=&\frac{1}{\sqrt{f_1f_5}}
\left(-hf_p^{-1}\d t^2+f_p[\d z+\frac{r_0^2}{r^2}\frac{\cosh\sigma_p\sinh
\sigma_p}{f_p}\d t]^2\right)\nn\\&&+\sqrt{f_1f_5}\left(\frac{1}{h}\d r^2+r^2
\d\Omega_3^2\right)\mtx{where}\\f_1&=&1+\frac{r_1^2}{r^2},
\;f_5=1+\frac{r_5^2}{r^2},\; h=1-\frac{r_0^2}{r^2},\;f_p=1+\frac{r_p^2}{r^2},
\;r_p^2=r_0^2\sinh\sigma_p^2.\nn
\eea 
In what follows we assume $r_p \!<\!<\!r_1,r_5$. Then, with the 
identifications \hfill\break 
$l^2\es r_1r_5,\;\rho^2\es \frac{R^2}{l^2}(r^2+r_0^2
\sinh^2\sigma_p),\; \tau\es \frac{l}{R}t$ and $\phi\es zR$, the metric \refs{5bs} 
approaches near the horizon ($r<\!< r_1,r_5$)
\be \label{ds1}
\d s^2=-f^2\d\tau^2+f^{-2}\d\rho^2+\rho^2(\d\phi-\frac{J}{2\rho^2}
\hbox{d}t)^2 +l^2\d\Omega_3^2\nonumber
\ee 
with
\be  f^2=\frac{(\rho^2-\rho_+^2)(\rho^2-\rho_-^2)}{l^2\rho^2}\mtx{and}J=2
\frac{\rho_+\rho_-}{l},\nn\ee which is recognised as that of the BTZ black 
hole \cite{BTZ,Carlip} with mass $8G\ell^2 M\es \rho_+^2+\rho_-^2$ 
and angular momentum $8G\ell J\es 2\rho_+ \rho_-$. Here $\ell$ is the 
inverse of the cosmological constant and $\rho_{+(-)}\es \frac{R}{\ell}r_0
\cosh \sigma_p(\sinh \sigma_p)$. 

\subsection{Decay Rate}The greybody factors for scalar fields in the BTZ 
black hole have been obtained in \cite{BSS1}. The resulting absorption 
cross section is given by   
\be\label{sigma2} \sigma_{abs}= {\cal{A}}_H\frac{| 
\Gamma(a_{{\sst +}}\!+1)\Gamma(a_{{\sst -}}\!+1)|^2}{|\Gamma
(a_{{\sst +}}\!+a_{{\sst -}}\!+1)|^2}\mtx{where}
a_{{\sst\pm}}=\frac{i\ell^2}{2(\rho_+\mp\rho_-)}\left(\omega\mp\frac{m}{\ell}
\right),\nn\ee  
or, when expressed in terms of the $5$-dimensional parameters, 
\be 
a_\pm=\frac{i(\omega_5\mp\frac{m}{R})\ell^2}{2r_0}e^{\pm\sigma_p},\nn
\ee 
hence reproducing the $5$-dimensional expression \cite{9609026} for both, 
neutral- and charged emission. The result \refs{sigma2} holds provided 
\be \label{adsc}\omega_5^2R^2-m^2<\!<\!1\mtx{and}[\omega_5\frac{R}{\ell}\cosh\sigma_p -\frac{m}{\ell}\sinh\sigma_p]r_0\frac{R}{\ell}<\!<\!1.
\ee 
It is interesting to compare \refs{adsc} to the $5$-dimensional 
string-theory condition \cite{9609026}
\be 
(\omega_5^2-(\frac{m}{R})^2)\ell^2<\!<\!1.
\ee 
For 'fat' black holes where $l>\!R$ the five dimensional 
condition is stronger whereas for $l<\!R$ the BTZ condition is stronger. 
Hence, for fat black holes the near horizon region still has a conformal 
field theory description for frequencies $1/\ell<\omega_5<1/R$ whereas 
the 'outside' geometry adds 'non-conformal' effects to the black hole. 
In the latter case the near horizon region is not big enough to capture 
all effects of the underlying conformal field theory. One has to include 
'outside' scattering to recover the conformal field theory description. 
Finally, we note that the equivalence continues to hold for charged 
emission where $\sigma$ has to be replaced by $\sigma'$ \cite{9609026}.

The above discussion can be repeated for the $4$-dimensional charged black 
hole and it is not hard to show that the $4$-dimensional greybody factors 
are equally well encoded in the near horizon BTZ geometry. Recently it has 
been argued \cite{9801125} that the $AdS_3$-structure holds for extremal 
black holes and $p$-brane geometries in various dimensions. We do not 
develop this issue further here but note that this is likely to extend the 
validity of the BTZ description. 

\section{Microscopic Theory}The quantisation of the BTZ-horizon was first 
proposed by Carlip \cite{Carlip} mapping the horizon degrees of freedom 
into a WZW-theory. This has lead to a microscopic computation of the entropy 
in terms of the horizon WZW-theory. An alternative approach \cite{9712251} 
concentrates on the asymptotic degrees of freedom. The two description should 
be equivalent as $3$-dimensional gravity is not dynamical. In the asymptotic 
description one uses the fact that the asymptotic isometries of the black 
hole metric form a Virarsoro algebra with central charge 
$c\es \frac{3\ell}{2G^{(3)}}$ \cite{Brown86}. 
Assuming that Cardy's formula \cite{Cardy86} applies\footnote{The 
applicability of Cardy's formula in its simplest form to the present 
situation has been questioned recently \cite{CC}, so that this result 
ought to be taken with a grain of salt} the statistical entropy for large 
black holes \cite{9712251} is identical with the geometrical 
entropy.

In order to describe the decay rate we need to know how the scalar field 
propagating in the gravitational background affects  the boundary conformal 
field theory. It turns out that its effect is to introduce an interaction of 
the form\footnote{The derivation of this result we be presented elsewhere }
\be 
S_{int}\propto\int \d x_+\d x_-\;I_+I_-
\;e^{i(\omega+m/\ell)x_++i(\omega-m/\ell)x_-},
\ee 
where $I_\pm$ are the left- and right moving Kac-Moody currents respectively. 
The transition amplitudes between various black hole states in the 
presence of a scalar field is therefore encoded in the set of correlation 
functions of Kac-Moody currents. To compute the decay rate of a highly 
excited black hole we sum over final states and thermally average over 
the initial states. The finite temperature two point function of the 
Kac-Moody currents is given by 
\be \< I_+(0)I_+(x_+)\>_T=2\left[\frac{\pi T}{\sinh(\pi T x_+)}\right]^2
\ee 
and similarly for the right moving part. 
The resulting emission rate precisely reproduces the semiclassical result 
\cite{BSS1}. 


\end{document}